# Numerical analysis of wave propagation across Solid–Fluid interface with Fluid–Structure interaction in circular tube

Tomohisa Kojima, Kazuaki Inaba

**Highlights**

- Numerical analysis of water confined in tube under axial impact

- Radial wave propagation shown to affect pressure distribution in tube

- Peak pressure exceeds value predicted by acoustic theory in specific region of tube

# Numerical Analysis of Wave Propagation across Solid–Fluid Interface with Fluid–Structure Interaction in Circular Tube

**Author names and affiliations**

**Tomohisa Kojima**[1]  (Corresponding author)

Graduate School of Science and Engineering, Tokyo Institute of Technology

2-12-1-I6-5, Ookayama, Meguro-ku, Tokyo, 152-8550 Japan

E-mail: kojima.31k@g.chuo-u.ac.jp

**Kazuaki Inaba**

School of Environment and Society, Tokyo Institute of Technology

**ABSTRACT**

Fluid–structure interaction (FSI) and wave propagation in engineering structures can cause severe damage to piping systems or fluid machines, inducing serious accidents. In these phenomena, the mechanism of structural damage depends on the wave propagation across the fluid–solid interface. Previous studies reported that disagreements between the induced pressure value on the solid–fluid movable interface and the value predicted by the classical one-dimensional theory arose from the effects of two-dimensional wave propagation. To address this problem, in this study, a two-dimensional axisymmetric simulation of wave propagation across the solid-fluid interface with FSI was conducted. The simulation was performed using ANSYS Autodyn with a Lagrangian solver for solids and Eulerian solver for water. The results showed that radial wave propagation caused by the dynamic effect of the tube and water's inertia affected the peak pressure on the solid–fluid interface. The peak pressure was attenuated near the tube wall because of the inertial effect of the tube and fluid expansion. By calculating the mean pressure and axial stress to compare the simulated peak pressure with that from one-dimensional acoustic theory, it was indicated that the transition region for transmitted pressure was located immediately after the solid–fluid interface. In this region, the transmitted peak pressure may exceed the value predicted by one-dimensional acoustic theory. The transition region was oriented in the axial direction from the interface. In addition, prediction of the transmitted peak pressure with one-dimensional acoustic theory was suggested via normal wave speed in the unconfined fluid from a safety engineering perspective, although the circumferential stress generated in the tube enclosing fluid can be sufficiently accurately predicted using the same theory with the Korteweg speed.

[1] Present address: Faculty of Science and Engineering, Chuo University
1-13-27, Kasuga, Bunkyo-ku, Tokyo, 112-8551 Japan

**1 INTRODUCTION**

Fluid–structure interactions (FSIs) can severely damage piping systems or fluid machines, inducing serious accidents, especially when they occur with wave propagation. Many studies have investigated FSI with wave propagation, including research on water hammers (Daude et al., 2018; Riedelmeier et al., 2017), cavitation (Chong and Kim, 2019; Triawan et al., 2015; Bergant et al., 2006), and structural responses to blast loading (Jin et al., 2015; Xue and Hutchinson, 2003). In these phenomena, the mechanism of structural damage depends on the wave propagation across the fluid–solid interface.

In the early stage, wave propagation crossing the fluid–solid interface was investigated by Taylor (1941). Taylor theoretically modeled the response of a rigid plate to exponentially decaying blast loading from a fluid in one dimensional (Swisdak, 1978). His model suggested that the impulse transmitted to the plate was reduced according to the reduction in plate mass. The effect of the reduction of plate mass in Taylor's model indicated that the inertial motion of the solid plate induced by the blast loading affected the interfacial and transmitted pressure on the fluid–solid interface. Many studies have further developed Taylor's model of FSI with a plate (Li et al., 2013; Wang et al., 2013; Liang et al., 2007). These studies are of interest for the construction of naval vessels and submarine components to investigate their resistance to explosive loading (Mouritz et al., 2001).

Some research has investigated wave propagation across fluid–solid interfaces using shock tubes. Many years after Taylor, Deshpande et al. verified Taylor's analysis experimentally using a shock tube (Deshpande et al., 2006). Schiffer et al. (2012) later developed Taylor's model analytically with different supporting conditions of a rigid plate and predicted the induced pressure and velocity of the plate. Schiffer and Tagarielli used the same experimental method as Dashpande et al. to analyze the dynamic response of composite plates (Schiffer and Tagarielli, 2014a, 2014b, 2015). Damazo and Shepherd (2017) experimentally observed the reflection of the gaseous detonation wave, reporting that the wave speed substantially exceeded the ideal theoretical value as well as the generation of a very high-pressure region near the gas–solid wall interface. Their research indicated the importance of clarifying interfacial phenomena concerning the propagation of stress or pressure waves across fluid–solid interfaces and the generation and transmission of interfacial pressure. More recently, Veilleux and Shepherd (2018, 2019) investigated the wave propagation with an air gap between solid and fluid both experimentally and numerically. They have been reported that the presence of a free surface and induced complex motion of the solid caused cavitation inception which results in the steepening of the pressure waves transmitted in the fluid (Veilleux and Shepherd, 2018, 2019).

In our previous studies, we have investigated wave propagation across a solid–fluid interface with FSI using a water-filled circular tube. These revealed that the classical one-dimensional theory of acoustic impedance could not predict the pressure induced on the solid–fluid interface with FSI (Kojima et al., 2017). In addition, it was implied that two-dimensional wave propagation by the radial expansion of the water-filled tube affected the interfacial pressure. In this study, we conducted a numerical simulation to investigate the two-dimensional wave propagation around the solid–fluid interface with FSI.

Many numerical methods have been developed and widely used to analyze FSI problems because they are of practical interest to many researchers for industrial applications. The hydrocode is one such simulation method (Zukas, 2004), suitable for analyzing phenomena in which the deformation and fracture of a medium are caused by stress or pressure wave propagation in the medium (Yagawa and Miyazaki, 2007). The formulation is made with strong coupling methods (Blom, 1998; Takashi and Hughes, 1992; Zhang and Hisada, 2001). The diffusion term, heat conduction, and viscosity are excluded from the hydrocode because the strong coupling method is intended to analyze transient phenomena in short periods with little effect from these terms.

This method is considered to arise from the HEMP code (Alder et al., 1964). Autodyn (Birnbaum et al., 1987) and LS-DYNA were developed from HEMP code as commercial hydrocodes.

In this study, we conducted numerical simulations using ANSYS Autodyn v.15.0. We chose this method because our target phenomenon of wave propagation across the interface of FSI is completed in a very short time. We conducted the numerical simulation with a two-dimensional axisymmetric model of wave propagation across the solid–fluid interface with FSI. The aim of this study is twofold. First, we intend to locate the transition region of a pressure wave in which the pressure cannot be predicted by classical one-dimensional theory, and second, we aim to propose an estimation method for the pressure in this transition region.

## 2 METHODS

### 2.1 Simulation model

We conducted a numerical study with ANSYS Autodyn v.15.0 and a 2D axially symmetric model of the problem. The problem geometry is presented in Fig. 1. This geometry was built with reference to the experiment conducted in our previous study (Kojima et al., 2017, 2015). The simulation model comprises two main parts: the cylindrical solid buffer and the water-filled circular tube. A steel projectile impact is used to generate a stress wave in the buffer. The generated stress wave propagates through the buffer and across the interface of the buffer and water in the tube. Thus, the stress wave in the solid is propagated to the water in the tube as a pressure wave. We used a Lagrangian solver for the solids and a Eulerian solver for water with the simulation model. In the experiment, the water was sealed with the buffer using an O-ring. In the simulation model, because of the complexity of modeling the O-ring seal, such as modeling the pre-stressed rubber and determination of the friction condition, water is stopped to spout out with the projection on the buffer in the gap between the buffer and tube wall. We used the projection of the buffer materials to avoid the occurrence of wave reflection between the buffer and the projection part. Over the simulation steps, the projection of the buffer and tube wall were not in contact.

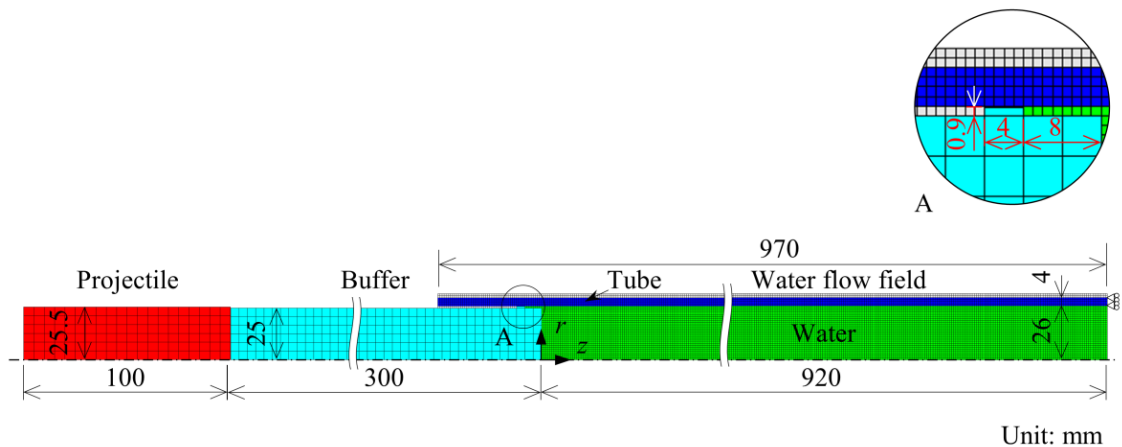

Fig. 1 Schematic of the simulation model used in the study

**2.2 Simulation conditions**

Table 1 presents the meshed condition of the model. The mesh sizes are decided considering Courant conditions because Autodyn is an explicit integration code (Robertson et al., 1994; Courant et al., 1928). In this simulation, the mesh conditions were set to be 4 meshes in the thickness direction of the tube wall. The simulation was also carried out under the conditions of finer to be 8 meshes in the thickness direction of the tube wall and the conditions of rougher to be 2 meshes. In each simulation result, the difference of the peak value of the hoop strain of the tube wall was within 10%. Therefore, it was considered that the simulation result was sufficiently converged. We adopted the condition of 4 meshes because the mesh would too rough in the condition of 2 meshes in the thickness direction of the tube wall.

Slip boundaries were applied to all interfaces between parts with no friction. The fixed condition was applied to the tube end, setting the velocity equal to 0 in the $z$-direction. The bottom of the water was closed with the end of the water flow field. Therefore, neither water nor pressure is transmitted through the bottom of the tube. Waves reaching the bottom were fully reflected. The material properties are given in Table 2. These material properties were taken from the literature (National Astronomical Observatory of Japan, 2016; Matsuka, 1984; Walley and Field, 1994). All the materials are modeled as linear-elastic. The linear equation used in this study is expressed with shock wave velocity $U_s$ and particle velocity $U_p$ as follows;

$$U_s = C_0 + s\, U_p \tag{1}$$

where $C_0$ is the bulk speed of sound, $s = dU_s/dU_p$. The equation of state is as follows;

$$p = K\left(\rho/\rho_0 - 1\right) \tag{2}$$

where $\rho_0$ is the density in the static state.

Originally, polycarbonate is a viscoelastic material, but the viscoelastic features should have little effect in the short time as in the present simulation (O'Connell and McKenna, 2002). Therefore, polycarbonate was also modeled as an elastic body in the simulation. The simulations were conducted below the elastic limits of all the materials. Aluminum and polycarbonate were used as the buffer materials with reference to the previous experiment. We set the velocity of the projectile as 1.40 m/s in the $z$-direction for the initial condition.

## 3  Simulation results

Figure 2 shows the stress and pressure distributions after the projectile impact during the first and second peak of the interfacial pressure. In Fig. 2, compressive pressure has a positive sign and thus tensile stress has a negative sign. The pressures below/above ±0.6 MPa are all shown in dark blue/red in Fig. 2. The range of the color scale was chosen to visualize pressure distribution in all frames. The generated compressive stress wave propagates in the buffer with continuous reflection at both edges of the buffer; when the wave reaches the buffer–water interface, it is transmitted to the water as a pressure wave. The transmitted wave indicates complex interactions between the water and the tube wall. There is a time-lag between the first peak pressure with the aluminum buffer and the polycarbonate buffer. It is due to the difference in sound speed between aluminum and polycarbonate.

The axial strain histories of the buffer both in the experiments and the simulation are presented in Fig. 3. The black lines in Figs. 3 to 6 are from the experimental results in the previous study (Kojima et al., 2017). These strain histories correspond to the incident stress wave profiles generated due to the collision of the projectile and the buffer. In the case with the aluminum buffer and the polycarbonate buffer, a difference of 50% and 35% was generated in the first peak value of the incident stress wave between the simulation and the experiment. Experiments confirmed repeated contact and separation when the projectile and buffer collided, and it seemed that the energy was lost when it was transferred from the projectile to the buffer during the collision. There is a limit to reproducing such complicated contact conditions on the simulation, then as a result, amplitudes of the incident stress waves generated by the collision have not been completely reproduced in this simulation. However, from Fig. 3, it appeared that the average transition of the incident stress wave was well reproduced in the simulation.

Figure 4 presents the velocity of the buffer-water interface both in the experiments and the simulation. In Fig. 4, 12% and 10% differences in the first peak value between simulations and experiments occurred in the case with the aluminum buffer and the polycarbonate buffer, respectively. It indicates the effect of the frictions of O-ring in the experiment.

Figure 5 plots the hoop strain histories of the water-filled tube at the outer tube wall. The amplitude of the numerical result is larger and steeper than that of the experimental result. As for the first peak value, the experimental results were 37% and 41% smaller than the simulated results with the aluminum buffer and the polycarbonate buffer, respectively. These differences would be mainly attributed to the difference of the incident stress wave between the experiment and the simulation (Fig. 3). As mentioned above, the amplitude of the incident stress wave could not be reproduced in the simulation in this study due to the imperfect collision of the projectile and the buffer in the experiment.

−0.6 MPa 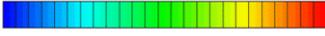 0.6 MPa

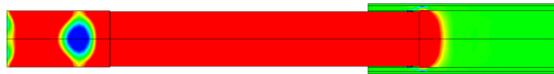

$t = 0.017$ ms, at the first peak of the interfacial pressure

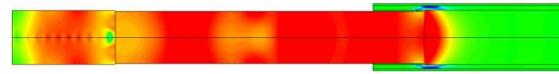

$t = 0.049$ ms, at the first peak of the interfacial pressure

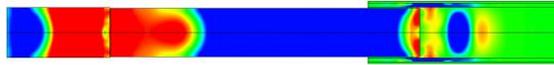

$t = 0.052$ ms

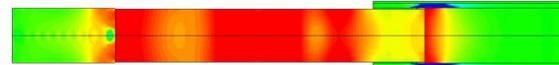

$t = 0.071$ ms

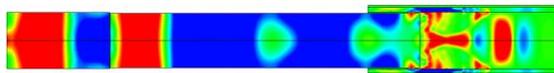

$t = 0.088$ ms

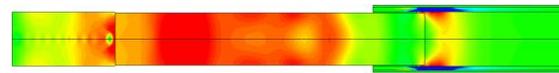

$t = 0.092$ ms

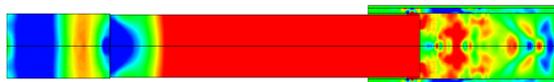

$t = 0.12$ ms

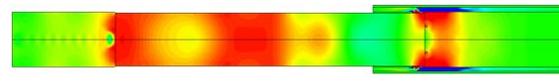

$t = 0.11$ ms

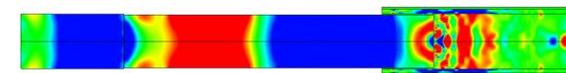

$t = 0.16$ ms (second peak of the interfacial pressure)

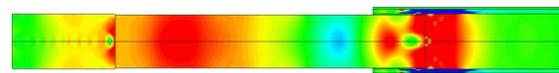

$t = 0.13$ ms (second peak of the interfacial pressure)

(a) (b)

Fig. 2 Stress and pressure distributions: (a) with the Al buffer, (b) with the PC buffer

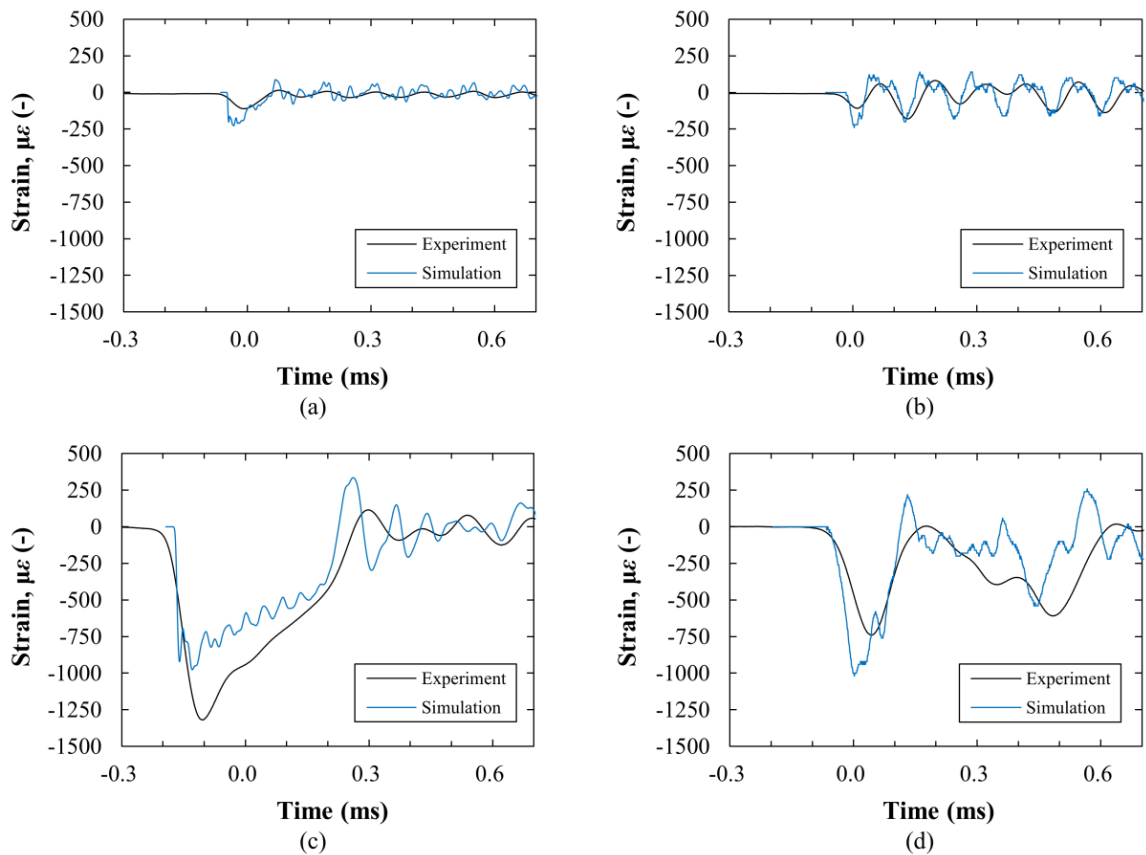

Fig. 3 Axial strain histories of the buffer: with the Al buffer (a) at $z = -280$ mm, (b) at $z = -80$ mm, with the PC buffer (c) at $z = -280$ mm, (d) at $z = -80$ mm

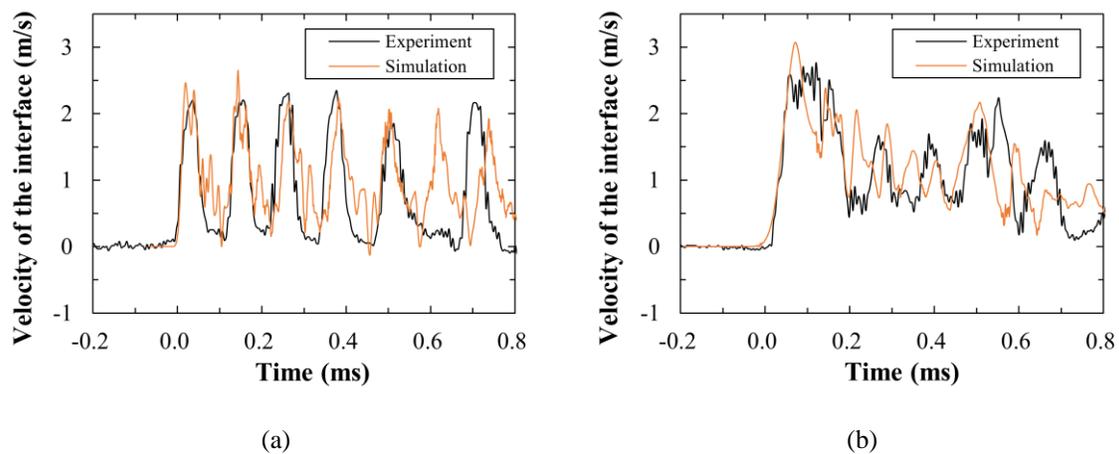

Fig. 4 Velocity of the buffer-water interface: (a) with the Al buffer, (b) with the PC buffer

There may be some other factors that caused the difference between the experiment and the simulation, such as friction due to the O-rings between the buffer and the tube wall, trapped air below the buffer, and three-dimensional effects in the experiments. However, it appeared that the reproducibility of the incident waves resulting from an imperfect collision of the projectile and buffer contributing by about 80% of the difference between the simulation and the experiments in the amplitude of the strain histories in Fig. 5. In general, the equation of state could cause the difference between the simulation and the experiment. In contrast, the phenomenon in this study could be reproduced by a linear equation of state which is used in this study because of the slow collision speed. However, despite the difference in the amplitude, the average transition of the strain history in the simulation was able to reproduce the experiment.

In Fig. 5, the hoop strain at each location in the tube-wall with the polycarbonate buffer is larger than the aluminum buffer. It may be attributed to the difference in the sound speed between aluminum and polycarbonate. Since the sound speed of aluminum is fast, even if the pressure in the water rises due to a stress wave transmitted into the water, the tensile wave generated by the free-end reflection at the end face of the buffer or projectile immediately returns and is transmitted into the water. As a result, the tube wall vibrates then cannot be fully pushed out by the pressure in water. On the other hand, because the sound speed of polycarbonate is slower, there is sufficient time for transmitted pressure waves to push and expand the tube wall without being affected by reflected tensile waves.

Figure 6 shows the pressure histories of water close to the buffer surface at the center of the tube ($r = 0$), as well as measured interfacial pressures from the experiment. The measuring location in the simulation was determined to be less than the maximum moving distance of the buffer. It is difficult to track the water pressure near the moving buffer tip because of the characteristics of the Eulerian coordinates. Therefore, the pressure measuring point was fixed in the simulation. If the gauge point in the simulation was less than $z = 2$ mm, the tip of the buffer would overlap with the gauge point, hindering pressure measurement in the water when it moved. In the experiment, measured values were filtered to eliminate noise with the natural pressure transducer frequency of 25 kHz. Therefore, the pressure fluctuation could be obtained in more detail with the numerical simulation. The pressure profile with the numerical simulation filtered with a 25 kHz low-pass filter was plotted in Fig. 6. The filtered pressure profile indicated a similar tendency with the experimental one.

As for the first peak value, the simulated results were larger than the experimental results. With the aluminum buffer and the polycarbonate buffer, the experimental results were 17% and 21% smaller for the pressure history at the center of the tube, respectively. These differences would be mainly attributed to the difference of the incident stress wave between the experiment and the simulation, as mentioned above.

Figure 4(a) indicates a pressure oscillation induced in water both in the experiment and the simulation. It should be due to stress waves transmitted into the water repeatedly while it reciprocates in the buffer in the z-direction. In other words, it was attributed to the axial vibration of the buffer due to the propagation of the stress wave. On the other hand, only the simulated pressure history oscillated in Fig. 4(b). This oscillation was similar to a radial-circumferential vibration mode of a circular pipe in asymmetry mode. Since the numerical simulation in this study is two-dimensional axisymmetric, the vibration mode of a higher order of asymmetry should not appear. However, as we discuss in the latter part in this section, it was confirmed that the pressure oscillated with the propagation of the pressure wave in the radial direction in the simulation. It is unclear how the asymmetric vibration mode which should exist affects the simulation result under the assumption of symmetry when the wave propagates radially in the two-dimensional axisymmetric analysis. Consequently, in the numerical simulation in this study, it could not completely reproduce the pressure oscillation in a circular tube filled with water. However, it would be said that the average transition of the pressure wave profile was reproduced.

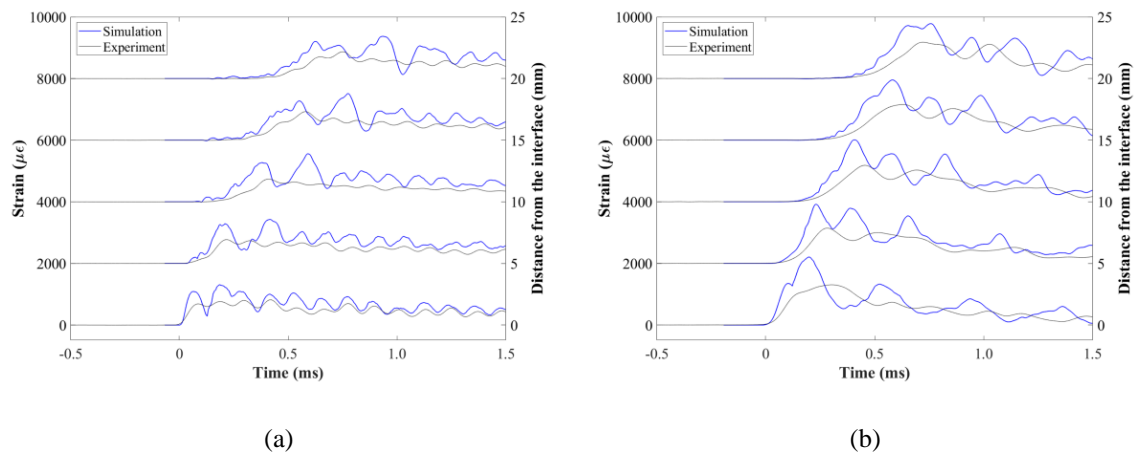

(a)          (b)

Fig. 5 Hoop strain histories of the water-filled tube at external tube wall: (a) with the Al buffer, (b) with the PC buffer

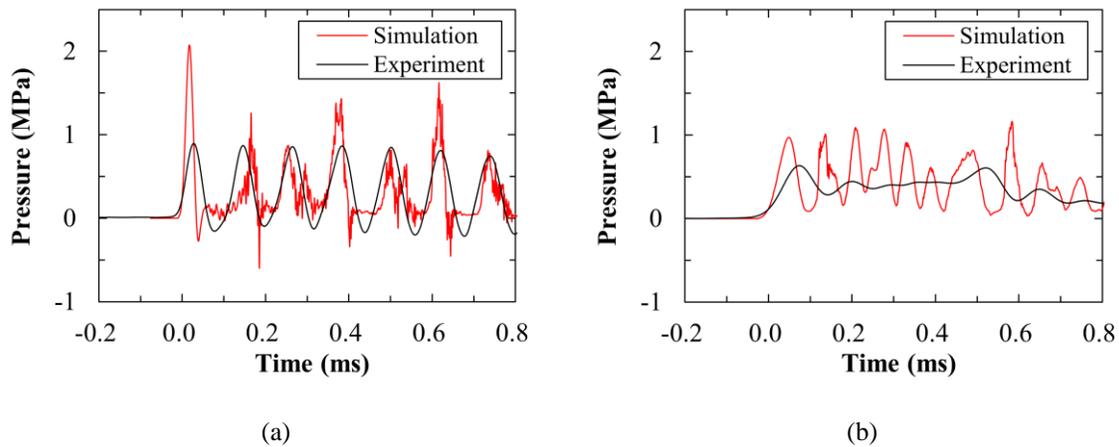

(a)                                                          (b)

Fig. 6 Pressure history in the water close to the buffer surface at $r$ = 0 mm: (a) with the Al buffer, (b) with the PC buffer

## 4 Effect of radial wave propagation

We investigated the effect of radial wave propagation with the obtained simulation result. Figure 7 shows the pressure distributions (compression is positive) and velocity vectors near the solid–fluid interface. The pressures below/above ±1 MPa are all shown in dark blue/red in Fig. 7. The range of the color scale was chosen to visualize pressure distribution in all frames. Near the tube wall, the transmitted pressure wave propagates to the tube wall and causes tube expansion (tensile stress) in the hoop direction after it crosses the buffer–water interface. As the transmitted pressure wave propagates farther in the $z$-direction, it is intricately distributed in the radial direction by interactions between the tube wall and water. Especially with the polycarbonate buffer, it seems that the pressure wave propagated in the radial direction in water due to the interactions between the tube wall and water after the first expansion of the tube wall. The pressure at the center with the polycarbonate buffer would reach the second peak due to the pressure wave propagation in the radial direction in the water. With the aluminum buffer, the pressure reached the second peak due to the transmission of the reflected stress wave from the buffer. Therefore, the mechanisms for reaching the second pressure peak at the center of the tube may different from both buffer materials.

To investigate the radial pressure distribution on the solid–fluid interface quantitatively, the pressure histories at $z$ = 2 mm and different radial locations are plotted in Fig. 8. The red lines in Fig. 8 ($r$ = 0 mm) is the same data with "simulation" lines in Fig. 6. In Fig. 8, the transmitted pressure peaks seem to be attenuated near the tube wall because of the tube expansion. The blue line in Fig. 8 is the water pressure near the inner tube wall, estimated using the hoop strain in the simulation recorded at the external tube wall via the following equation proposed by Tijsseling (2007):

$$p - p_{\text{out}} = E_{\text{tube}} \frac{e}{R}\left(1 + \frac{1}{2}\frac{e}{R}\right)(\varepsilon_\varphi - \varepsilon_{\varphi,\text{out}}) \tag{3}$$

where $\varepsilon_{\varphi,\text{out}} = -(1 - v)p_{\text{out}}/E$ is the hoop strain of the external tube wall when $p = p_{\text{out}}$. Equation (1) is derived assuming a quasi-static relation between the stress in the tube and pressure in the water. However, in Fig. 8, the estimated pressures do not agree with the inner pressure histories in water. This reveals that the transmitted peak pressure is strongly affected by the dynamic effect of the tube and water's inertia.

Here, we compared the simulation results with the one-dimensional acoustic theory to clarify the effect of radial wave propagation. According to the acoustic theory, the amplitude of transmitted stress crossing the interface of the different solid media can be estimated with the following equation (Meyers, 1994):

$$\sigma_{\text{T}} = \frac{2\frac{Z_2}{A_2}}{\frac{Z_1}{A_2} + \frac{Z_2}{A_1}} \sigma_{\text{I}} \tag{4}$$

where $\sigma_{\text{T}}$ is the transmitted stress, $\sigma_{\text{I}}$ is the incident stress, $Z$ is the acoustic impedance of the medium ($Z = \rho c$ where $\rho$ is density and $c$ is sound speed), and $A$ is the cross-sectional area. The index numbers distinguish the two solid media. As it stands, the acoustic theory for wave transmission and reflection is derived only for solid–solid or fluid–fluid discontinuous interfaces. Here, we applied Eq. (4) for a solid–fluid interface to predict the interfacial pressure.

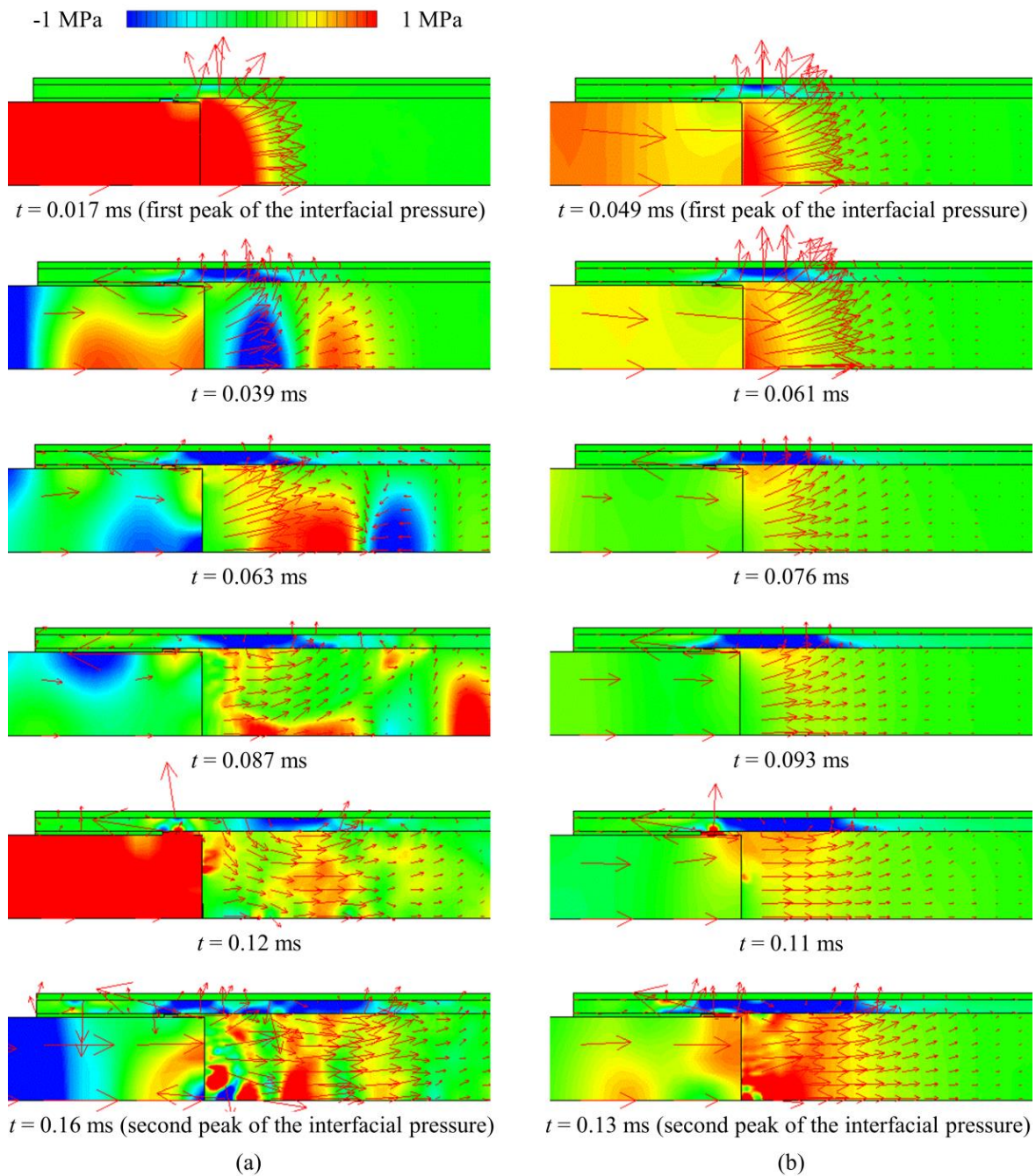

Fig. 7 Pressure distributions and velocity vectors: (a) with the Al buffer, (b) with the PC buffer

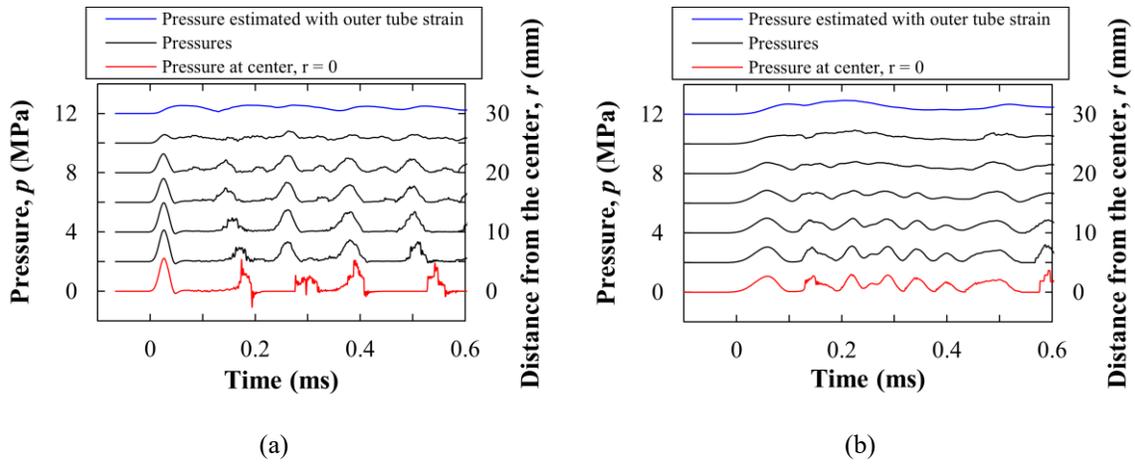

Fig. 8 Pressure histories at $z = 2$ mm: (a) with the Al buffer, (b) with the PC buffer

To obtain the acoustic impedance of water, we used the Korteweg speed of sound in the calculation as the sound speed in the water. The Korteweg speed is the wave speed when the wave propagates axially in an elastic tube interacting with the tube wall (Korteweg, 1878). The Korteweg speed $c_K$ is defined as follows:

$$c_K = \frac{c_w}{\sqrt{1 + 2R_0 K/E_{tube}e}} \qquad (5)$$

where $c_w$ is the sound speed of the unconfined water, $R_0$ is a representative value of the tube radius, $K$ is the bulk modulus of the fluid, $E_{tube}$ is the elastic modulus of the tube, and $e$ is tube thickness. For the buffer, the longitudinal wave speed $c_B$ was calculated assuming isotropy (Eqs. (6) and (7)). The obtained sound speed and acoustic impedance are listed in Table 3.

$$c_B = \sqrt{\frac{E}{\rho}} \qquad (6)$$

$$E = \frac{9KG}{3K+G} \qquad (7)$$

We plotted the first peak pressure of water at every 5 mm in the $r$-axis in Fig. 9. The dashed lines in Fig. 9 represent the transmitted pressures according to the one-dimensional acoustic theory derived with Eqs. (4)–(7). Fig. 9 reveals that the peak pressure in water is radially distributed at each location in the axial direction. This may be caused by the rarefaction wave generated by tube expansion. When the pressure wave expands the tube wall, the rarefaction wave is generated near the tube wall because of the tube motion. With the Al buffer, because the wave speed in Al is much faster than that in water, wave transmission from the buffer to water may be completed in a shorter period compared to the rarefaction wave propagation. Therefore, the distribution of peak pressure is larger in Fig. 9(a). However, because the wave speed is slower with the PC buffer, the rarefaction wave affects the center of the tube and decreases the peak pressure (Fig. 9(b)). The distributed peak values gradually converge to a value slightly below the acoustic theoretical value as the pressure wave propagates away from the interface.

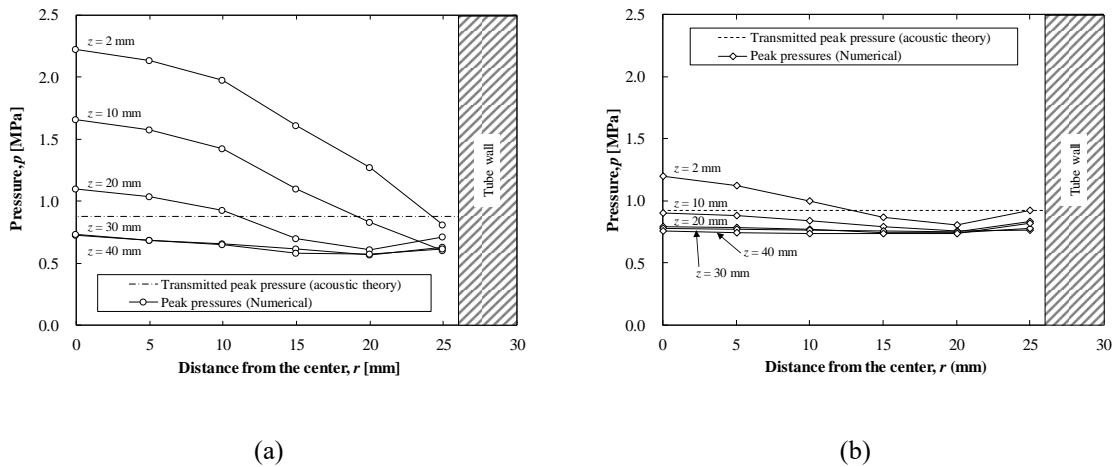

(a)         (b)

Fig. 9 Radial distribution of the peak pressure: (a) with the Al buffer, (b) with the PC buffer

Furthermore, we converted the distributed peak pressure and axial stress to one-dimensional mean pressure/axial stress by adopting the cross-sectional area mean of the peaks at each location in the $z$-direction. Figure 10 illustrates our use of the area mean for calculating the pressure or axial stress. Using Eqs. (8) and (9), we estimated the pressure or axial stress at each location along the $z$-direction.

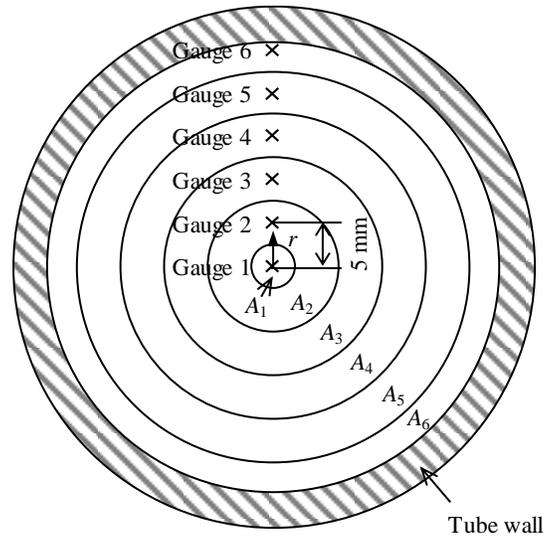

Fig. 10 Illustration of area division for estimation of mean pressure or axial stress

$$\bar{p} = \frac{\sum_{i=1}^{6} p_{i\,peak} \cdot A_i}{\sum_{i=1}^{6} A_i} \tag{8}$$

$$\begin{cases} A_1 = \pi(5/2)^2 \\ A_{i=2\sim5} = \pi\{(r_i + 5/2)^2 - (r_i - 5/2)^2\} \\ A_6 = \pi\{r_6^2 - (r_6 - 5/2)^2\} \text{ (for the buffer)} \\ A_6 = \pi\{(r_6 + 1)^2 - (r_6 - 5/2)^2\} \text{ (for water)} \end{cases} \tag{9}$$

where *i* is the numbering of gauge points in the radial direction, as shown in Fig. 10.

Figure 11 shows the shifting of the estimated peak pressure or axial stress along the *z*-direction. The incident axial stress is estimated using the following equation of impact theory (Eq. (10)). Equation 8 describes the amplitude of the axial stress wave generated by the collision of two cylindrical solids, assuming continuities of pressure or axial stress and particle velocity across the interface with the wave equation (Goldsmith 2002; Hayashi and Tanaka 1988).

$$\sigma_2 = \frac{A_1 E_1 E_2 V_1}{A_2 E_2 c_1 + A_1 E_1 c_2} \tag{10}$$

where $V_1$ is the impact velocity. In this study, we used $V_1 = \sqrt{2gh}$, and applied the projectile to medium 1 (index number 1) in Eq. (10) and the buffer to medium 2 (index number 2).

In Fig. 11, the estimated axial stress starts to decline before $z = 0$ because of the reflected tensile wave from the interface (at $z = 0$). After the wave propagates across the interface, the transmitted peak pressure converges at $z = 30$ and 10 mm with the aluminum and polycarbonate buffers, respectively. The convergence is judged when the difference of the pressure peak value becomes less than 10% with the value at a position away from the interface. The region up to the point of convergence from the interface can be considered the transition region of the transmitted pressure wave. In this transition region, the transmitted peak pressure can exceed the estimated value of one-dimensional acoustic theory with the Korteweg speed ($c_K = 434$ m/s). Especially with the Al buffer, the peak pressure is 61% larger than the estimated value at $z = 2$ mm (Fig. 11(a)). Hence in some cases, it may be dangerous to predict the pressure using the Korteweg speed for designing a piping structure near the solid–fluid interface. As an index, the length of the transition region can be written as $z_T/D = 0.58$ and 0.19 for the aluminum and polycarbonate buffers, respectively, where $z_T$ is the length of the transition region in the axial direction and $D$ is the inner diameter of the tube.

The length of the transition region should be differed by the intensity of the generated pressure distribution, the material properties of the solid and fluid, and the shape parameters such as the thickness of the tube wall. Specifically, it is considered to depend on each parameter contained in Eqs. (4) to (6). In this study, the effect of the sound speed of the buffer material has been examined. Besides, the present result is a result that holds only within the range of elastic deformation of the solid. Further study is needed to clarify how the parameters in Eqs. (4) to (6) affect the length of the transition region.

The disagreement of the simulation result with theory in the transition region indicates the transition of the tube expansion. In the transition region, the transmitted pressure wave begins interacting with the tube wall to expand it. Therefore, the deformation state of the tube wall is in transition in the region. We plot the transmitted peak pressures according to the acoustic theory with the wave speed of unconfined water ($c_w = 1,491$ m/s) in Fig. 11 as dashed lines. With the wave speed of unconfined water, the simulation results remain below the theoretical value after the interface, including the transition region. Consequently, in the design of a piping structure, the transmitted peak pressure should be estimated using the wave speed of the unconfined fluid in the transition region near the solid–fluid interface from the perspective of safety engineering. A similar transition region may exist in cases such as pipe ends, expanding or shrinking pipes, unrestrained valves, bends, and branches. These are known as cases of junction coupling in liquid–pipe interaction (Tijsseling 1996). In addition, orifices may also have transition regions. In this way, the prediction of the pressure peak value in the transition region requires further study with different materials, shapes, and sizes of both tubes and fluids.

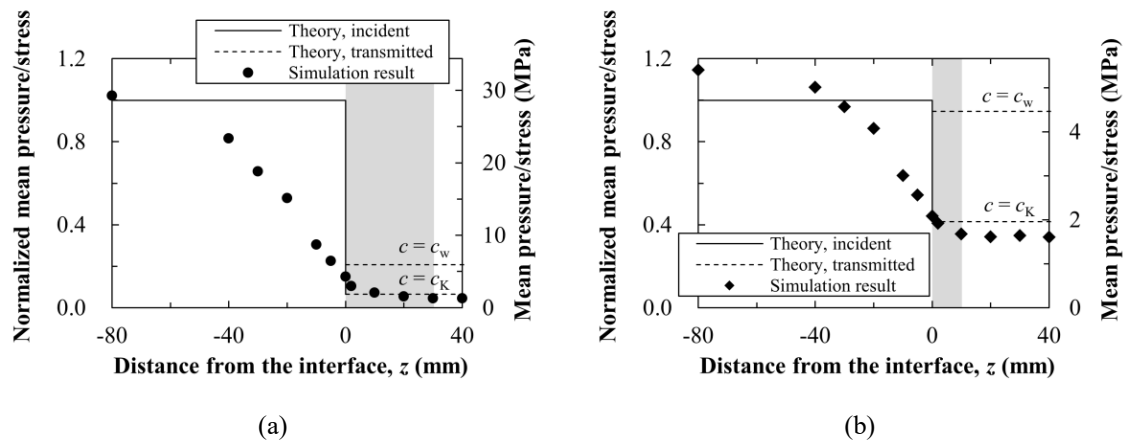

(a) (b)

Fig. 11 Shifting of the mean peak pressure/axial stress: (a) with the Al buffer, (b) with the PC buffer

In addition, the mean pressure is converted to the circumferential stress generated in the pipe, then compared with the simulation results, experimental results, and acoustic theory (Fig. 12). From Fig. 12, the tube stresses in the circumferential direction predicted from the mean pressures in the transitional regions are shown to be greater than those predicted by acoustic theory using the Korteweg velocity, especially with the Al buffer (Fig. 12 (a)). However, it is understood that the circumferential stress generated in the tube is slightly lower than the theoretically predicted value in both the experimental result and the simulation result. This is because the pressure exerted on the pipe is reduced by the pressure distribution, as shown in Fig. 9. Therefore, when the stress wave is transmitted through the solid–fluid interface with FSI and propagated as a pressure wave in the fluid, it can be said that the one-dimensional acoustic theory using the Korteweg velocity can predict the circumferential stress generated in the tube enclosing the fluid with sufficient accuracy, regardless of the transitional region of the pressure.

In this study, we focused on a case in which a stress wave is transmitted from the solid to fluid in an elastic tube. The transition region is considered to appear due to the interaction between the elastic tube wall and fluid. Therefore, the concept of transition region proposed in this study would not be applicable to the case of wave transmission from fluid to solid. It would not be applicable also to the case of wave reflection on the interface. Further study is needed about the wave transition across the solid-fluid/fluid-solid interface.

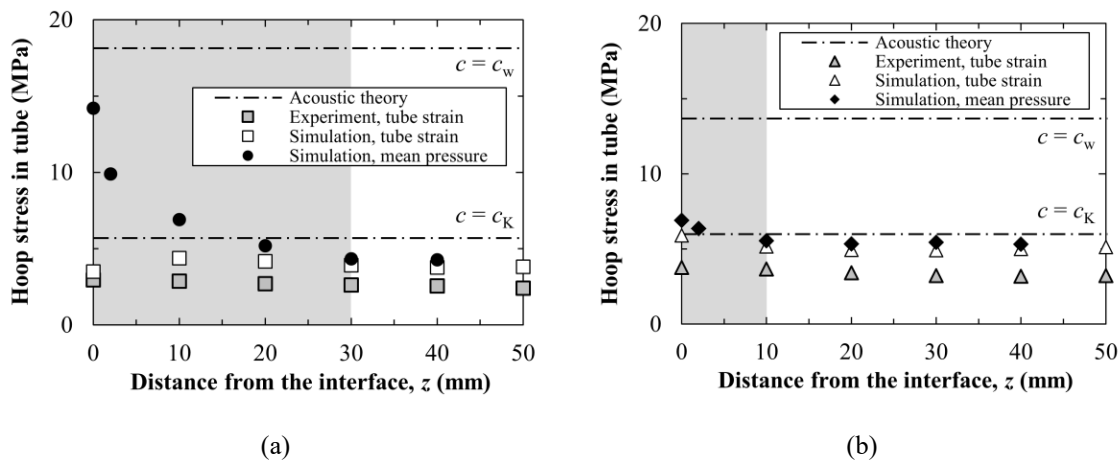

Fig. 12 Stress peak in the tube in hoop direction: (a) with the Al buffer, (b) with the PC buffer

**5 Conclusion**

A numerical study was conducted with wave propagation across the solid–fluid interface with FSI. The problem was modeled in two-dimensional axial symmetry with a cylindrical solid buffer and a water-filled circular tube. It was revealed that the mechanisms causing the peak pressure on the solid–fluid interface differed depending on the buffer material, varying for aluminum and polycarbonate. With the aluminum buffer, the pressure on the interface peaked when the stress wave propagating axially in the buffer reached the interface. On the other hand, with the polycarbonate buffer, the interfacial pressure oscillated, and its peaks were caused by radial pressure wave propagation. It was revealed that the transmitted peak pressure was strongly affected by the dynamic effects of the tube and water's inertia.

The peak pressure was attenuated near the tube wall because of the inertia effect of the tube and fluid expansion. The averaged pressure distribution after the transmission gradually converged to that predicted by one-dimensional acoustic theory as the pressure wave propagated away from the interface. By calculating the mean pressure to compare the simulated peak pressure with that predicted by one-dimensional acoustic theory, it was indicated that a transition region for the transmitted pressure existed immediately after the solid–fluid interface. In this region, the transmitted peak pressure exceeded the value predicted by one-dimensional acoustic theory for the aluminum buffer. The length of the transition region $z_T$ was expressed as a function of the inner diameter of the tube $D$, extending to the point $z_T/D = 0.58$ and $0.19$ for the aluminum and polycarbonate buffers, respectively in the axial direction from the interface. In this region, the transmitted peak pressure should be estimated with the one-dimensional acoustic theory using the normal wave speed in the unconfined fluid from a safety engineering perspective. However, the circumferential stress generated in the tube enclosing the fluid can be predicted with sufficient accuracy using the same theory and the Korteweg speed.

Further study is needed with more various materials and scale of the diameter of the buffer, tube, and fluid, also with an unmovable solid buffer for application of this study to the engineering of structures intended for transporting or storing moving fluids, such as piping and reactors.

**Acknowledgment**

We are grateful to Prof. Kikuo Kishimoto for fruitful discussion. Funding: This work was supported by JSPS KAKENHI [grant numbers JP26709001, JP18K13662]. Declarations of interest: none.

Table 1. Meshed condition for the simulations

| Parts | Projectile | Buffer | Tube | Water flow field |
| --- | --- | --- | --- | --- |
| Grid ($i \times j$) | $20 \times 6$ | $75 \times 6$ | $970 \times 4$ | $970 \times 32$ |
| Total grids | 120 | 450 | 3880 | 31040 |

Table 2. Material properties for the simulations

| Material | Aluminum (Al) | Polycarbonate (PC) | Water | Stainless steel |
|---|---|---|---|---|
| Applied part | Buffer | Buffer, Tube | Water | Projectile |
| Equation of state | Linear | Linear | Linear | Linear |
| Constitutive equation | Elastic | Elastic | - | Elastic |
| Density, $\rho$ [kg/m$^3$] | 2,680 | 1,200 | 1,000 | 8,000 |
| Bulk modulus, $K$ [GPa] | 75.5 | 2.72 | 2.22 | 194 |
| Shear modulus, $G$ [GPa] | 26.9 | 1.00 | - | 73.0 |

Table 3. Calculated sound speed and acoustic impedance for each material

|  | Al | PC | Water in the PC tube | Stainless steel |
|---|---|---|---|---|
| $c$ [m/s] | 5,171 | 1,492 | 434* | 4,932 |
| $Z$ [$10^6$ kg/m$^2$·s] | 14.0 | 1.79 | 0.434* | 39.5 |

*with the Korteweg speed

Figure Legends

Fig. 1 Schematic of the simulation model used in the study

Fig. 2 Stress and pressure distributions: (a) with the Al buffer, (b) with the PC buffer

Fig. 3 Axial strain histories of the buffer: with the Al buffer (a) at $z$ = -280 mm, (b) at $z$ = -80 mm, with the PC buffer (c) at $z$ = -280 mm, (d) at $z$ = -80 mm

Fig. 4 Velocity of the buffer-water interface: (a) with the Al buffer, (b) with the PC buffer

Fig. 5 Hoop strain histories of the water-filled tube at external tube wall: (a) with the Al buffer, (b) with the PC buffer

Fig. 6 Pressure profile in water at $z$ = 2 mm, $r$ = 0 mm: (a) with the Al buffer, (b) with the PC buffer

Fig. 7 Pressure distributions and velocity vectors: (a) with the Al buffer, (b) with the PC buffer

Fig. 8 Pressure histories at $z$ = 2 mm: (a) with the Al buffer, (b) with the PC buffer

Fig. 9 Radial distribution of the peak pressure: (a) with the Al buffer, (b) with the PC buffer

Fig. 10 Illustration of area division for estimation of mean pressure or axial stress

Fig. 11 Shifting of the mean peak pressure/axial stress: (a) with the Al buffer, (b) with the PC buffer

Fig. 12 Stress peak in the tube in hoop direction: (a) with the Al buffer, (b) with the PC buffer